\documentclass[twocolumn,aps,prl,showpacs,preprintnumbers,amsmath,amssymb]{revtex4}
\usepackage{graphicx}
\usepackage{color}
\usepackage{dcolumn}
\usepackage{bm}
\newcommand{\gev}{\, \text{GeV}}
\newcommand{\kev}{\, \text{keV}}

\newcommand{\bra}{\langle}
\newcommand{\ket}{\rangle}

\begin{document}

\title{Three-Nucleon Low-Energy Constants from the Consistency of\\
  Interactions and Currents in Chiral Effective Field Theory} 
\author{Doron Gazit}
\affiliation{Institute for Nuclear Theory, University of Washington, 
Box 351550, 98195 Seattle, Washington, USA}
\author{Sofia Quaglioni}
\affiliation{Lawrence Livermore National Laboratory, P.O. Box 808, L-414, Livermore, CA 94551, USA}
\author{Petr Navr\'{a}til}
\affiliation{Lawrence Livermore National Laboratory, P.O. Box 808, L-414, Livermore, CA 94551, USA}
\date{\today}

\begin{abstract}
The chiral low-energy constants $c_D$ and $c_E$ are constrained by means of accurate {\em ab initio} calculations of the $A\!=\!3$ binding energies and, for the first time, of the triton $\beta$ decay. We demonstrate that these low-energy observables 
allow a robust determination of the two undetermined constants, 
a result of the surprising fact that the determination of $c_D$ depends weakly on the short range correlations in the wave functions.
These two- plus three-nucleon interactions, originating in chiral effective field theory and constrained by properties of the $A\!=\!2$ system and the present determination of $c_D$ and $c_E$, are successful in predicting properties of the $A\!=\!3$, and 4 systems.  
\end{abstract}


\maketitle
The fundamental connection between nuclear forces and the underlying theory of quantum chromodynamics (QCD) remains one of the greatest contemporary theoretical challenges, due to the non-perturbative character of QCD in the low-energy regime relevant to nuclear phenomena.
However, the last two decades of theoretical developments provide us with a bridge to overcome this obstacle, in the form of chiral perturbation theory ($\chi$PT)~\cite{chiPT-WE90}. 
The $\chi$PT Lagrangian, constructed by
integrating out degrees of freedom of the order of
$\Lambda_\chi\sim1 \gev$ and higher (nucleons and pions are thus the only explicit degrees of freedom), is an effective Lagrangian of QCD at low energies. As such, it retains all conjectured symmetry principles, particularly the approximate chiral symmetry, of the underlying theory.
Furthermore, it can be organized in terms of a perturbative expansion 
in positive powers of  
$Q/\Lambda_\chi$ where $Q$ is the generic momentum in the nuclear process or the pion mass~\cite{chiPT-WE90}. 
Though the subject of an ongoing debate about its validity~\cite{Weinberg-counting-problems,potentials-review}, the naive extension of this expansion to non-perturbative phenomena provides a practical interface with existing many-body techniques, and clearly  holds a significant value for the study of the properties of QCD at low energy and its chiral symmetry.

The chiral symmetry dictates the operator structure of each term of the effective Lagrangian, whereas the coupling constants (not fixed by the symmetry) carry all the information on the integrated-out degrees of freedom.
A theoretical evaluation of these coefficients,
or low-energy constants (LECs), is equivalent to solving QCD at low-energy. 
Recent lattice QCD calculations have allowed a theoretical estimate of LECs of single- and two-nucleon diagrams~\cite{Lattice}, while LECs of diagrams involving more than two nucleons are out of the reach of current computational resources.  
Alternatively, the undetermined constants can be constrained by low-energy experiments.  

\begin{figure}[t]
\rotatebox{0}{\resizebox{5.5cm}{!}{
\includegraphics[clip=true,viewport=2.5cm 18.5cm 20.5cm 27.5cm]{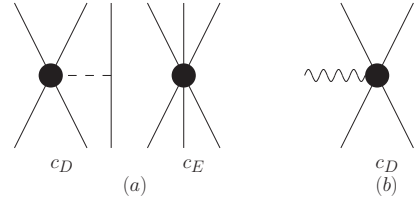}} }
\caption{Contact and one-pion exchange plus contact interaction ($a$), and contact MEC ($b$) terms of $\chi$PT. }
\label{Fig:diagram}
\end{figure}
The strength
of $\chi$PT is that the chiral expansion is used to derive both nuclear potentials and currents from the same Lagrangian. Therefore, the electroweak currents in nuclei (which determine reaction rates in processes involving external probes) and the strong interaction dynamics ($\pi N$ scattering, the $NN$ interaction, the $NNN$ interaction, etc.) are all 
based on the same theoretical grounds and rooted in the low-energy limits of QCD. 
In particular, $\chi$PT predicts, along with the $NN$ interaction at the leading order (LO), a three-nucleon ($NNN$) interaction at the next-to-next-to-leading order or N$^2$LO~\cite{N2LO-NNN,nd-constrain}, and even a four-nucleon force at the fourth order (N$^3$LO)~\cite{N3LO-NNNN}. At the same time, the LO nuclear current consists of (the standard) single-nucleon terms, while two-body currents, also known as meson-exchange currents (MEC), make their first appearance at N$^2$LO~\cite{MEC}.  
Up to N$^3$LO  both the $NNN$ potential and the current are fully constrained by the parameters defining the $NN$ interaction, with the exception of  two ``new" LECs, $c_D$ and $c_E$.  The latter, $c_E$, appears only in the potential as 
the strength of the $NNN$ contact term [see Fig.~\ref{Fig:diagram}\,$(a)$]. On the other hand, $c_D$ manifests itself both in the contact term part of the  $NN$-$\pi$-$N$  three-nucleon interaction of Fig.~\ref{Fig:diagram}\,($a$) and in the  two-nucleon contact vertex with an external probe of the exchange currents [see Fig.~\ref{Fig:diagram}\,$(b)$]. 

First attempts to determine $c_D$ and $ c_E$ have used the triton binding energy (b.e.) alongside an additional strong observable, either the $nd$ doublet scattering length~\cite{nd-constrain}, or the $^4$He b.e.~\cite{7Li}. However, this led to a substantial uncertainty in the values of the LECs, due to correlations between the $^3$H b.e. and these observables, known respectively as the Phillips and Tjon lines. The fine-tuning of these observables is very sensitive to the structure of the adopted $NNN$ force. Hence small variations of the cutoff, different regularization schemes, missing terms of the interaction, etc., tend to produce large swings in the extracted  of $c_D$ and $c_E$. A different approach was adopted in Ref.~\cite{p-shell-constrain}. There, a preferred choice for the two LEC's was obtained by complementing the constraint on the $A\!=\!3$ b.e.\ with a sensitivity study on the radius of $^4$He 
and on various properties of $p$-shell nuclei. The same interaction was then successfully used to predict the $^4$He total photo-absorption cross section~\cite{4He-cs}.

The need for a complemental determination of these LECs is two-fold. First, it would be desirable to perform such a determination within the $A\le 3$ systems, so to suppress any additional many body contribution. 
Second, despite recent progress \cite{N3LO-missing-pieces}, the $NNN$ potential has been fully worked out only up to N$^2$LO, leaving inconsistency in the calculation of the wave functions when combined with the NN force at N$^3$LO. Clearly, it would be preferable to adopt an observable with minimal dependence on the short-range part of the wave function.
In this respect, the relation (mandated by the chiral symmetry of QCD) between electroweak processes and $NNN$-force effects offers venues to achieve these goals. 
This relation was established in the context of effective field theory~\cite{pi-production,Phillips,Nakamura}, and manifests itself  in $\chi$PT via the appearance of $c_D$ in both the $NN$-$\pi$-$N$ diagram of  Fig.~\ref{Fig:diagram}\,($a$) and the one in Fig.~\ref{Fig:diagram}\,($b$).
 In particular, G{\aa}rdestig and Phillips~\cite{Phillips}, 
 suggested the triton beta-decay as one of the electroweak processes that could be used as input to fix the strength of the $NNN$ force. It is the purpose of this {\it{Letter}} to undertake this task and show that by using the triton half life, as well as the $A\!=\!3$ b.e., one can constrain the two undetermined LECs within the three-nucleon sector by means of fully converged {\em ab initio} calculations. We demonstrate that this determination is robust. The resulting chiral Lagrangian 
predicts, without any free parameters, various  $A\!=\!3$, and 4 properties. 

The triton is an unstable nucleus, which undergoes $\beta$-decay with a ``comparative" half-life  of $(fT_{1/2})_t\!=\!(1129.6\pm 3)$ s
\cite{exp-half-life}.
This quantity can be used to extract an empirical value for 
$\langle E_1^A\rangle\!=\!|\langle^3{\rm He}||E_1^A||^3{\rm H}\rangle|$~\cite{Simpson,Schiavilla},
the reduced matrix element of the $J\!=\!1$ electric multiple of the axial vector current, through 
  \begin{equation}
 (fT_{1/2})_t = \frac{K/G_V^2}{(1-\delta_c)+3\pi\,\frac{f_A}{f_V}\,\langle E_1^A\rangle^2}\,.
 \end{equation}
Here,  $K\!=\!2\pi^3\ln{ 2}/m_e^5$ (with $m_e$ the electron mass), $G_V$
is the weak interaction vector coupling constant (such that
$K/G_V^2\!=\!6146.6\pm0.6 \text{s}$~\cite{Hardy}), $f_A/f_V\!=\!1.00529$~\cite{Schiavilla} accounts for the small difference in the statistical rate function between
vector and axial-vector transitions, and
$\delta_c\!=\!0.13\%$~\cite{Schiavilla} is a small
correction to the reduced matrix element of the Fermi operator,
calculated between the $A\!=\!3$ wave functions (which is 1 for  this
specific case) due to isospin-breaking in the nuclear interaction. 
One can use these values to extract
$\langle E_1^A\rangle|_{emp}\!=\!0.6848\pm0.0011$. 

The weak axial current adopted in this work is the N{\"{o}}ther current built from the axial symmetry of
the chiral Lagrangian up to order N$^3$LO~\cite{MEC}. At LO this current 
consists of the standard single-nucleon part, which at low momentum transfer is proportional to the Gamow-Teller (GT) operator,
$E_1^A|_{\rm LO}\!=\!  i\,g_A(6\pi)^{-1/2}\sum_{i=1}^A \sigma_i \tau^{+}_i\,,$
where $\sigma_i$, $\tau_i^+$ are spin and isospin-raising operators of the $i$th nucleons, and $g_A\!=\!1.2695 \pm 0.0029$ is the axial constant~\cite{PDBook}. For this reason, the quantity $\sqrt{3\pi}g_A^{-1} \langle E_1^A\rangle|_{emp}$ is often referred to as ``experimental", or ``empirical", GT. 

Corrections to the single-nucleon current appear at  
N$^2$LO in the form of MEC and relativistic terms.
The MEC are formed by a one-(charged)-pion exchange, and a contact term.
While the relativistic corrections are negligible for the triton half life, the MEC have a substantial influence on this $\beta$-decay rate. This is a reflection of the fact that $E_1^A$ is a chirally unprotected operator \cite{chiral_filter}. Moreover, the strength of the MEC contact term, usually denoted by $\hat d_R$, is related to $c_D$ through:
\begin{equation}
\hat{d}_R \equiv \frac{M_N}{\Lambda_\chi g_A} c_D +\frac{1}{3} M_N
(c_3 + 2c_4) +\frac{1}{6}.
\end{equation}
Here, $M_N$ is the nucleon mass, and $c_3$ and $c_4$ are LECs of the dimension-two $\pi N$ Lagrangian, already part of the chiral $NN$ potential at NLO.
Therefore, one can use 
$\langle E_1^A\rangle|_{emp}$ as second constraint for the determination of $c_D$ and $c_E$.

\begin{figure}[t]
\rotatebox{0}{\resizebox{6.5cm}{!}{
\includegraphics{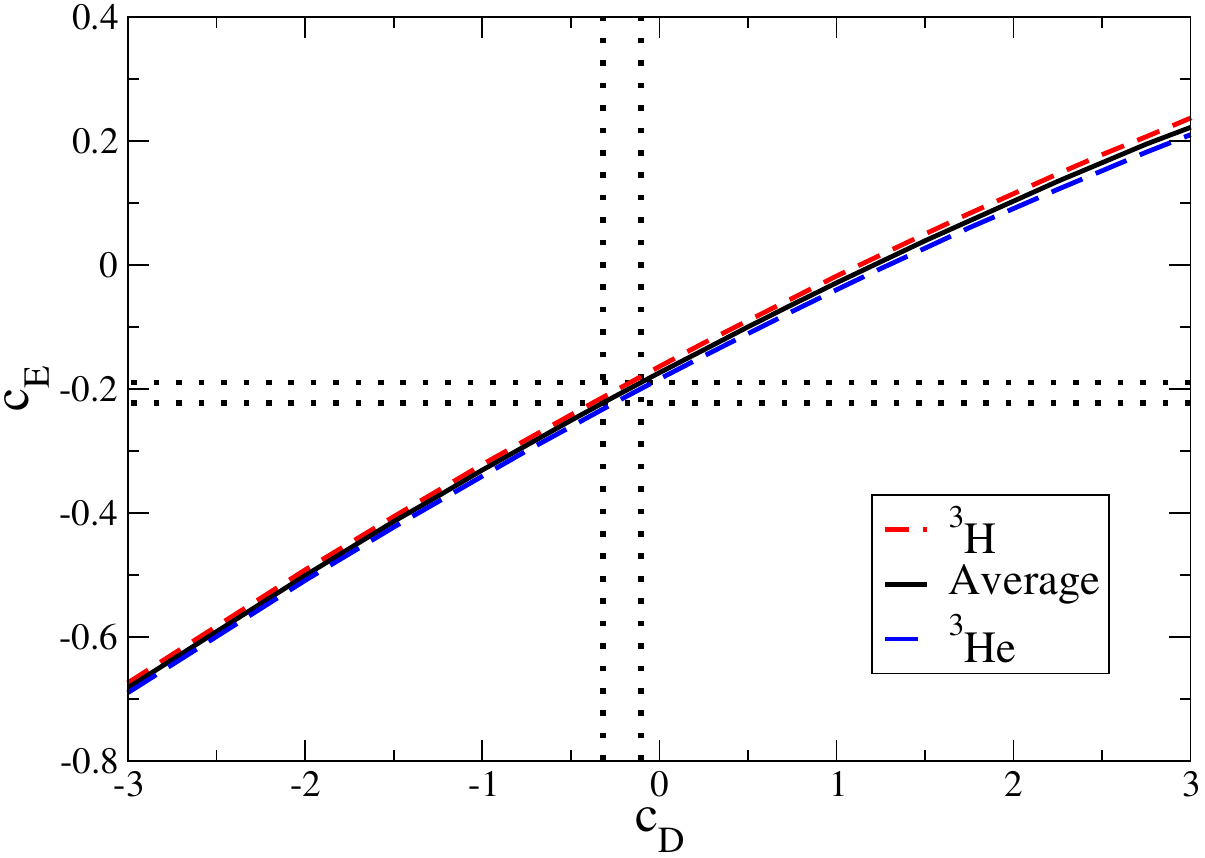}} }
\caption{(Color online.) $c_D$-$c_E$ trajectories fitted to reproduce $^3$H and $^3$He experimental
  b.e. The dotted lines show the region for which $|1-\bra E_1^A\ket_{th}/\bra E_1^A\ket_{emp}|$ is within the $\pm 0.54\%$ error-bars.}
\label{fig:calibrate}
\end{figure}
Following the $c_D$-$c_E$ trajectory which reproduces, on average, the $A\!=\!3$ b.e., as discussed in Ref~\cite{p-shell-constrain}, 
here, we (i) calculate the $^3$H and $^3$He g.s. wave functions by solving the Schr\"odinger equation for three nucleons interacting via the $\chi$PT $NN$ potential at N$^3$LO of Ref.~\cite{N3LO-NN} and the $NNN$ interaction at N$^2$LO~\cite{N2LO-NNN} in the local form of Refs.~\cite{p-shell-constrain,4He-cs,N2LO-local,footnote} ; (ii) determine for which $c_D$ values along the trajectory the calculated 
reduced matrix element of the $E_1^A$ operator (at N$^3$LO) reproduces 
$\bra E_1^A\ket_{emp}$.
 
The present calculations are performed in the framework of the no-core shell model (NCSM) approach~\cite{7Li,p-shell-constrain,N2LO-local,NCSM}. 
This method looks for the eigenvectors of the Hamiltonian in the form of expansions over a complete set of harmonic oscillator (HO) basis states up to a maximum excitation of $N_{\rm max}\hbar\Omega$ above the minimum energy configuration, where $\Omega$ is the HO parameter. 
Thanks to the large model-space size adopted  
($N_{\rm max}\!=\!40$), $A\!=\!3$ b.e.\  
and  reduced matrix element of $E_1^A$ are converged  to less than $0.05\%$.
Note that the same regulator $F_\Lambda(q^2)\!=\!\exp(-q^4/\Lambda^4)$ is used for both $NNN$ terms of the interaction and MEC, a process resulting in a local chiral $NNN$ force (for relevant parameters and definitions see Ref.~\cite{N2LO-local}). The $A\!=\!3,4$ calculations of Ref.~\cite{N2LO-local} were later confirmed by the results of Ref.~\cite{Pisa}, providing a benchmark for the local chiral $NNN$ force. The MEC utilized in this work were validated against those of Park {\em et al}.~\cite{MEC}. Finally, we tested the implementation of the MEC within the NCSM approach by reproducing (within 0.1\%) the AV18 \cite{AV18} results for $\langle E^1_A\rangle$ obtained 
using the effective-interaction hyper-spherical harmonics approach~\cite{MEC}.

\begin{figure}[t]
\rotatebox{0}{\resizebox{6.8cm}{!}{
\includegraphics{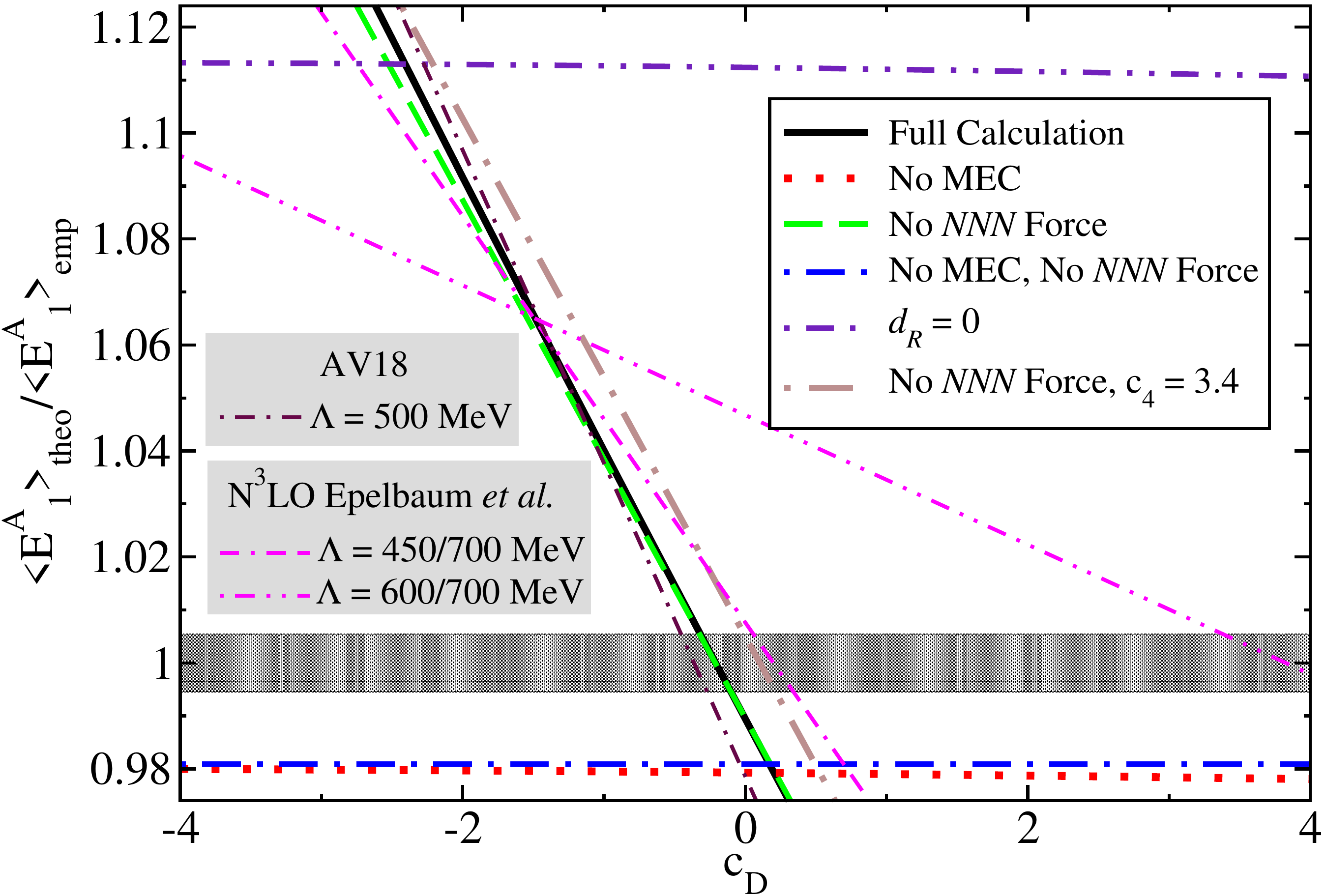}} }
\caption{(Color online.)  The ratio $\bra E_1^A\ket_{th}/\bra E_1^A\ket_{emp}$ 
using the N$^3$LO $NN$ potential~\cite{N3LO-NN} with and without the local N$^2$LO $NNN$ interaction~\cite{N2LO-local},  and the axial current with and without MEC ($c_D,c_E$ are varied along the averaged trajectory of Fig.~\ref{fig:calibrate}). The shaded area is twice the experimental uncertainty. Also shown: results (without $NNN$ force) for the phenomenological AV18 potential (with $\Lambda=$ 500 MeV imposed in the current), and for the N$^3$LO $NN$ potential of Epelbaum et al.~\cite{Epelbaum-NN} (with $\Lambda=450, 600$ MeV, and a 700 MeV spectral-function cutoff in the two-pion exchange).} 
\label{Fig:checks}
\end{figure}
\begin{table*}[t]
\begin{ruledtabular}
\caption{Calculated $^3$H, $^3$He and $^4$He g.s. energies (in MeV) and point-proton radii (in fm), obtained using the N$^3$LO $NN$ potential~\cite{N3LO-NN} with and without the local N$^2$LO $NNN$ interaction~\cite{N2LO-local} with $c_D\!=\!-0.2$ and $c_E\!=\!-0.205$, compared to experiment.}
\begin{tabular}{lcccccc}
\label{predictions}
&  \multicolumn{2}{c}{$^3$H} & \multicolumn{2}{c}{$^3$He} & \multicolumn{2}{c}{$^4$He}\\\cline{2-3}\cline{4-5}\cline{6-7}\\[-2mm]
&  $E_{\rm g.s.}$ & $\langle r^2_p\rangle^{1/2}$ & $E_{\rm g.s.}$ & $\langle r^2_p\rangle^{1/2}$ & $E_{\rm g.s.}$  & $\langle r^2_p\rangle^{1/2}$ \\ [0.7mm]
\hline\\[-3mm]
$NN$& $-$7.852(4) & 1.651(5) & $-$7.124(4) & 1.847(5) & $-$25.39(1)& 1.515(2)\phantom{1~\cite{alpharadius}} \\
$NN\!+\!NNN$&$-$8.473(4) & 1.605(5) & $-$7.727(4) & 1.786(5) & $-$28.50(2) & 1.461(2)\phantom{1~\cite{alpharadius}}\\
Expt. &$-$8.482$\phantom{(5)}$ & 1.60$\phantom{8(5)}$ & $-$7.718\phantom{(1)} & 1.77\phantom{7(1)} & $-$28.296\phantom{()} & 1.467(13)~\cite{alpharadius}
\end{tabular}
\end{ruledtabular}
\end{table*}

The theory to empirical value ratio for the $E_1^A$ reduced matrix element 
along the averaged constraint of Fig.~\ref{fig:calibrate} (which reproduces the $A\!=\!3$ b.e. to about $10\kev$~\cite{p-shell-constrain})
is presented in Fig.~\ref{Fig:checks}. The 1.08\% tolerance band highlighted by the shaded area (obtained by doubling the error bar) is mainly due to the uncertainties on $\bra E_1^A\ket |_{emp}$ and $g_A$. Besides the full calculation, which appears as a solid line, we report also the results of several tests, aimed to analyze the sensitivity of the triton half life to $NNN$ force and/or MEC.  

First we note the fundamental importance of the axial two-body currents in reaching agreement with experiment.  
By suppressing the MEC,  in the whole investigated $c_D$-$c_E$ range, the calculations under-predict  $\bra E_1^A\ket|_{emp}$ by about 2\%. The same, almost constant, behavior is found when adding to the single-nucleon current only the long-range one-pion-exchange term of the MEC, which corresponds to  setting $\hat d_R\!=\!0$. In this case, the theoretical results over-predict $\langle E_1^A\rangle|_{emp}$ by $\sim$11\%. Only when adding the contact part of the MEC, which is related to the short range weak correlations of axial character, can the half-life reach its experimental value. In particular, we find  that agreement within $\pm 0.54\%$ of the empirical value is obtained for
$-0.3\!\leq\! c_D\!\leq\! -0.1$. The corresponding $c_E$ values lie in the range $[-0.220, -0.189]$.   These results are summarized 
by the dotted lines in Fig.~\ref{fig:calibrate}.

In a similar spirit, we now study the effect of the suppression of the $NNN$ force. If we try to calibrate $c_D$ to reproduce the measured half-life, we obtain a curve in close agreement with the results of the full calculation~\cite{Phillips} (for completeness we show also the curve corresponding to the suppression of both MEC and $NNN$ force).  Moreover, a quantitatively similar $c_D$ dependence of $\bra E_1^A \ket$ can be obtained using $A\!=\!3$ wave functions produced by the phenomenological AV18 $NN$ potential (without $NNN$ force).  It is therefore clear that the half life of triton presents a very weak sensitivity to the $NNN$ force,  
 and hence to the strength of the spin-orbit interaction.  Thanks to this feature, which might be unique to $s$-shell nuclei, we are confident that the determination of $c_D$ and $c_E$ obtained in this way is robust.  
Incidentally, the weak dependence of the half life of triton upon the $NNN$ force can also explain the success of recent calculations done in a hybrid approach, coined EFT*~\cite{MEC}. 

As the values of the $c_3$ and $c_4$ LECs are somewhat uncertain (see, e.g., Ref.~\cite{7Li}), 
it is important to assess to which extent they would influence the determination of $c_D$ from the triton half life. While very sensitive to the smallest change in $c_3$,  the N$^3$LO fit of the $NN$ data of Ref.~\cite{N3LO-NN} does not deteriorate dramatically for $3.4\gev^{-1}\!\!\le\!c_4\!\le\!5.4\gev^{-1}$~\cite{Machleidt-private}. 
Fig.~\ref{Fig:checks} shows calculations (without $NNN$ force) carried out  by setting $c_4$ to $3.4\gev^{-1}$ ($\pi N$ value~\cite{c4piN}) in the axial current, while the $A\!=\!3$ wave functions are still obtained from the N$^3$LO $NN$ potential of Ref.~\cite{N3LO-NN} (where, in GeV$^{-1}$, $c_3\!=\!-3.2$ and $c_4\!=\!5.4$).   
We find that the use of 
the lower $c_4$ value produces a shift  ($\sim$0.3) towards more positive $c_D$ values.

Finally, Fig.~\ref{Fig:checks} shows two additional curves (without $NNN$ force) obtained using the N$^3$LO $NN$ potential of Ref.~\cite{Epelbaum-NN} with $\Lambda=450$ and $600$ MeV (700 MeV two-pion exchange spectral-function cutoff) and the parameters defining it (particularly, in $\gev^{-1}$, $c_3=-3.4$ and $c_4=3.4$). As one could expect, the extracted value of the $NNN$ LEC $c_D$ depends on the choice of the cutoff. However, we observe that the $\Lambda=450/700$ MeV potential gives comparable results as the N$^3$LO $NN$ potential of Ref.~\cite{N3LO-NN} and AV18, indicating that the determination of $c_D$ depends mainly on the MEC cutoff and weakly on the cutoff imposed in the nuclear potential. 

With this calibration of $c_D$ and $c_E$, for this potential,
in principle, any other calculation 
is a prediction of $\chi$PT. In Table~\ref{predictions} we present a collection of $A\!=\!3$ and 4 data, obtained with and without inclusion of the $NNN$ force for $c_D\!=\!-0.2$ ($c_E\!=\!-0.205$), a choice in the middle of the constrained interval. Besides triton and $^3$He g.s.\ energies, which are by construction within few keV from experiment, the $NN\!+\!NNN$ results for the $^4$He 
are in good agreement with measurement. Note that $\alpha$ particle g.s.\  energy and point-proton radii change minimally with respect to variations of  $c_D$ in the interval $[-0.3,-0.1]$, and the results at the extremes are both within the numerical uncertainties quoted in Table~\ref{predictions}. This result is not inconsistent with the study 
of Ref.~\cite{p-shell-constrain}, which showed preference for $c_D\!\sim\!-1$,
since for $p$-shell nuclei one expects the (neglected) higher-order $NNN$ force terms to affect, probably through a shift, the value of $c_D$ \cite{N3LO-missing-pieces}.

Summarizing, we have used 
the $A\!=\!3$ b.e.\ and the half-life of triton to constrain the undetermined N$^3$LO $\chi$PT parameters of the $NNN$ force. We have demonstrated the robustness of the constraint on $c_D$ by showing the weak sensitivity of the $\langle E_1^A\rangle$ matrix element with respect to the $NNN$ force. In particular, we find $-0.3\!\le\! c_D\!\le\!-0.1$, and, correspondingly, $-0.220\!\le\! c_E\!\le\!- 0.189$. The latter is expected to change due to N$^3$LO terms of the $NNN$ interaction, which were not included thus far.
In conclusion,  we have identified a clear path towards determining the $NNN$ force that, once the $NN$ interaction will be pinned down, will pave the way to 
parameter-free predictions of QCD in the consistent approach provided by $\chi$PT.

We thank S.\ Coon and U.\ van Kolck for valuable discussions.
Numerical calculations have been partly performed at the LLNL LC facilities.
Prepared in part by LLNL under Contract DE-AC52-07NA27344. S.Q.\ and
P.N.\ acknowledge support from the U.\ S.\ DOE/SC/NP (Work Proposal
No.\ SCW0498), and from the U.\ S.\ Department of Energy Grant
DE-FC02-07ER41457. D.G.\ acknowledges support from U.\ S.\ DOE Grant DE-FG02-00ER41132.  


\end{document}